
\magnification=1200
\hoffset=-.1in
\voffset=-.2in

\def\ll{\langle}
\def\rr{\rangle}
\def\diag{{\rm diag}}

\vsize=7.5in
\hsize=5.6in
\tolerance 10000

\baselineskip 12pt plus 1pt minus 1pt
\pageno=0
\centerline{\bf ELECTROMAGNETIC FIELDS OF A MASSLESS}
\smallskip
\centerline{{\bf PARTICLE AND THE EIKONAL}\footnote{*}{This
work is supported in part by funds
provided by the U. S. Department of Energy (D.O.E.) under contract
\#DE-AC02-76ER03069.\smallskip}}
\vskip 24pt
\centerline{R. Jackiw, D. Kabat\footnote{$^\dagger$}{National Science
Foundation Graduate
Fellow.\smallskip} and M. Ortiz\footnote{$^\ddagger$}{SERC fellow}}
\vskip 12pt
\centerline{\it Center for Theoretical Physics}
\centerline{\it Laboratory for Nuclear Science}
\centerline{\it and Department of Physics}
\centerline{\it Massachusetts Institute of Technology}
\centerline{\it Cambridge, Massachusetts\ \ 02139\ \ \ U.S.A.}
\vskip 1.5in
\centerline{Submitted to: {\it Physics Letters B\/}}
\vfill
\centerline{Typeset in $\TeX$ by Roger L. Gilson}
\vskip -12pt
\noindent CTP\#2033\hfill November 1991
\eject
\baselineskip 24pt plus 2pt minus 2pt
\centerline{\bf ABSTRACT}
\medskip
Electromagnetic fields of a massless charged particle are described by a gauge
potential that is almost everywhere a pure gauge.  Solution of
quantum mechanical wave equations in the presence of such
fields is therefore immediate and leads to a new derivation of
the quantum electrodynamical
eikonal approximation.  The electromagnetic action in the eikonal limit is
localized on a contour in a two-dimensional Minkowski subspace of
four-dimensional space-time. The exact $S$-matrix of this reduced theory
coincides with the eikonal approximation.
\vfill
\eject
Some years ago, 't~Hooft derived the eikonal approximation for the
two-particle scattering amplitude arising from their gravitational
interaction.$^1$  However, he did not sum eikonal graphs (graviton exchange
graphs with particle propagators approximated by neglecting virtual graviton
momenta compared to particle momenta).  Rather he solved quantum mechanical
equations for one particle moving in the gravitational field of the other,
which is taken massless; {\it i.e.\/} external gravity is described by the
Aichelburg--Sexl metric.$^2$  Recently, this problem has been re-examined by
H. and E.~Verlinde,$^3$ and their work sparked our interest in the subject.

We present here a similar approach to the electrodynamical eikonal.  We first
determine the electromagnetic fields of a massless charged particle.  This is
a classical physics textbook problem; because we could not locate its solution
in the literature  we present the formulas, which are interesting and elegant.
In particular, we find that the gauge potentials are pure
gauges, but {\it different\/} in {\it different\/} regions of space-time.  The
solution of a quantum mechanical wave equation is therefore obtained by
exponentiating the gauge function, and the scattering amplitude is the quantum
electrodynamical eikonal.

To find the electromagnetic fields of a charged ($e$) massless point particle,
we begin with those arising from a massive particle moving with constant
velocity $v$ along the $+z$-axis, and then set $v$ to the velocity of light,
which we scale to unity.  Our metric tensor is $\diag(1,-1,-1,-1)$, and
four-vectors $x^\mu$ are indexed by Greek letters from the middle of the
alphabet.  In the two-dimensional ($t,z$) Minkowski subspace, we index
by initial Greek letters: $\alpha,\beta,\ldots$; also we use light cone
components $\pm\equiv{1\over \sqrt{2}}(0\pm3)$.  Transverse Euclidean vectors
$r^i_\perp$ carry a Latin index.

Either by integrating the Li\'enard--Wiechert potentials or by Lorentz-boosting
the static Coulomb potential, one finds that the conserved source current
$$j^\mu_{\rm v} = en^\mu_{\rm v} \,\delta^2({\bf r}_\perp) \delta(z-vt),\qquad
n^\mu_{\rm v}=(1,0,0,v)\eqno(1)$$
gives rise to the 4-vector potential
$$A^\mu_{\rm v} = {e\over 4\pi} \ {n^\mu_{\rm v} \over R_{\rm v}}\ \ ,\qquad
R^2_{\rm v} = (z-vt)^2 + (1-v^2) r^2_\perp\eqno(2)$$
Since the $v=1$ limit of (2) is somewhat delicate, see below, we record
first the field strengths.
$$\eqalign{E^z_{\rm v} &= {e(z-vt)\over 4\pi}\ {1-v^2\over R^3_{\rm v}}\
 \ , \cr
B^z_{\rm v} &= 0 \ \ ,\cr}\hskip .5in
\eqalign{
E^i_{\rm v} &= {er^i\over 4\pi}\ {1-v^2\over R^3_{\rm v}}\cr
B^i_{\rm v} &= - {e\epsilon^{ij} r^j_\perp v\over 4\pi}\ {1-v^2\over R^3_{\rm
v}
}\cr}
\eqno(3)$$
It is straightforward to establish (for example by Fourier transforming with
respect to $z$) that
$$\lim\limits_{v\to 1} {1-v^2\over R^3_{\rm v}} = {2\over r^2_\perp}
\delta(t-z)
\eqno(4)$$
Thus the fields of a massless charged particle, moving in the $+z$-direction,
are
$$\eqalign{ E^z &= 0 \ \ ,\cr
B^z &= 0\ \ , \cr}\hskip .5in
\eqalign{
E^i &= {er^i_\perp\over 2\pi r^2_\perp} \delta(t-z) \cr
B^i &= - {e\epsilon^{ij} r^j_\perp\over 2\pi
r^2_\perp} \delta(t-z) \cr}\eqno(5)$$
and one may verify that they satisfy the Maxwell equations with the source
current
$$j^\mu = en^\mu \,\delta^2({\bf r}_\perp) \delta(t-z)\ \ ,
\qquad n^\mu = (1,0,0,1)
\eqno(6)$$

The potentials that give rise to (5) can be chosen as
$$A^0_{\rm I} = A^z_{\rm I} = 0\ \ ,\qquad {\bf A}^\perp_{\rm I} = - {e\over
2\pi} \theta(t-z) {\nabla}\ln \mu r_\perp\eqno(7)$$  
where $\mu$ is an irrelevant parameter, setting the scale of the logarithm.

For completeness, we derive the limiting forms (7) from $A^\mu_{\rm v}$ in (2).
First note that$^4$
$$\lim\limits_{v\to1} {1\over R_{\rm v}} = - \ln \mu^2 r^2_\perp \,\delta(t-z)
+
{1\over |t-z|} \eqno(8\hbox{a})$$
a formula that is consistent with (4) and the relation ${\partial\over\partial
r^2_\perp} R^{-1}_{\rm v} = - {1\over 2} (1-v^2) R^{-3}_{\rm v}$.  Here $\mu$
is
 a
necessary regulator.  The validity of (8a) is seen after a Fourier
transformation with respect to $z$; then it states
$$\lim\limits_{v\to1} K_0 \left(\omega r_\perp \sqrt{1-v^2}\,\right) = -
\ln \mu r_\perp + \lim\limits_{\epsilon\to0} \int^\infty_\epsilon {dz\over z}
\cos z\omega\eqno(8\hbox{b})$$
Both the modified Bessel function $K_0$ and the integral give rise to
singularities in the limit.  Upon identifying $\mu$ with
$\sqrt{1-v^2}\,\big/2\epsilon$, (8a) follows.  We conclude therefore that
$$\lim\limits_{v\to1} A^\mu_{\rm v} = - {en^\mu\over 4\pi} \left( \ln \mu^2
r^2_\perp \,\delta(t-z) - {1\over |t-z|}\right) \eqno(8\hbox{c})$$
The last term does not contribute to the fields; it is a pure gauge,
albeit its gauge function is singular at $t=z$.  Thus the gauge potentials can
be taken as
$$A^0_{\rm II} = A^z_{\rm II} = - {e\over 2\pi} \ln\mu r_\perp \delta(t-z) \ \
,\qquad {\bf A}^\perp_{\rm II} = 0 \eqno(9)$$
They differ from (7),
but of course they are related by a gauge transformation.
$$A^\mu_{\rm I} = A^\mu_{\rm II} + \partial^\mu\Omega\ \ ,\qquad \Omega =
{e\over 2\pi} \ln \mu r_\perp \,\theta(t-z)\eqno(10)$$

Both gauges will be useful in the following. Expression (7) shows that
$A^\mu_{\rm I}$ vanishes at $t<z$ and is a pure gauge for $t>z$, hence it is a
singular pure gauge.  $A^\mu_{\rm
II}$ has the advantage that the gauge potentials vanish in the transverse
directions, being confined to the two-dimensional $(t,z$) Minkowski space.

Quantum mechanical wave equations are conveniently analyzed in gauge I,
Eq.~(7).   When the electromagnetic interaction of a particle with charge $e'$
enters through the covariant derivative $\partial_\mu + ie'A_\mu$, its wave
function does not interact with $A_\mu$ at early times $t<z$,
$$\psi_< = \psi_0 \eqno(11\hbox{a})$$
while at late times $t>z$,
$$\psi_> = \exp -\left( i {ee'\over 4\pi} \ln \mu^2 r^2_\perp\right)
{\psi_0}'
\eqno(11\hbox{b})$$
Here $\psi_0$ and ${\psi_0}'$ are free wave functions, related by a continuity
requirement at $x^-=0$.
$$\psi_> \big|_{x^-=0} = \psi_<\big|_{x^-=0} \eqno(11\hbox{c})$$
Upon taking the initial wave function to be a plane wave,
$$\psi_0 = e^{-ip\cdot x} \eqno(12)$$
the derivation of the scattering amplitude follows 't~Hooft's to his
expression with $-Gs$ replaced by $ee'/4\pi\equiv\alpha$,
as was already noted by him.$^1$ One finds for distinguishable, spinless
particles
$$f(s,t) = {\Gamma(1+i\alpha)\over 4\pi i\mu^2\Gamma (-i\alpha)} \left(
{4\mu^2\over -t}\right)^{1+i\alpha} = {\alpha\over \pi t} \
{\Gamma(1+i\alpha)\over \Gamma(1-i\alpha)} \exp\left( i \alpha \ln
{4\mu^2\over -t}\right) \eqno(13)$$
where $s$ and $t$ are the Mandelstam variables.  ('t~Hooft scales $\mu^2$ to
unity and apparently omits the prefactor $1/i$.)

That this (and 't~Hooft's gravity
result) is exactly the eikonal formula, is seen
by recalling the standard expression of the latter.$^5$
$$\eqalign{f_{\rm eikonal}(s,t) &= i\int {d^2
b\over (2 \pi)^2} e^{i{\bf q}_\perp\cdot {\bf b}} \left[ 1 -
\exp \left( -iee'\int {d^2 k_\perp\over(2\pi)^2} \ {e^{i{\bf k}\cdot {\bf
b}}\over k^2_\perp + \mu^2}\right) \right] \cr
t &= - q^2_\perp \cr}\eqno(14)$$
Here ${1\over k^2_\perp+\mu^2}$ is the photon propagator at
$k_+k_-=0$, but with an
infrared regulating ``mass'' $\mu$.  The $k_\perp$ integral
leaves ${1\over 2\pi}K_0(\mu b)$, which for small $\mu$ is replaced by
$-{1\over 2\pi} \ln {e^\gamma\over 2} \mu b$.
  Absorbing $e^\gamma/2$ in $\mu$, and performing the $b$
integral reproduces (13).
$$f_{\rm eikonal} (s,t) = {\Gamma(1+i\alpha) \over
4\pi i \mu^2\Gamma(-i\alpha)} \left( {4\mu^2\over -t}
\right)^{1+i\alpha} \eqno(15)$$

Next we show that high energy eikonal electrodynamics can be given an action
formulation where the action is localized on a contour in the $(t,z)$ plane.
This is the analog in the present context of what has been done for Einstein
gravity$^3$.  For particles moving rapidly along the $z$-axis, the light
cone components of the source current are functions of only one light
cone coordinate, as seen in (6).
$$\eqalign{j_+(x) &= j_+(x^+,{\bf r}_\perp)\cr
           j_-(x)&= j_-(x^-,{\bf r}_\perp)\cr
           j^i(x) &= 0\cr}\eqno(16)$$
For such a source current, $E^z$ and $B^z$ vanish.  We may therefore take
${\bf A}^\perp = 0$ and $A_{\pm} = \partial_{\pm} \Omega$, as in gauge II,
Eq. (9).  Also we work in Landau gauge $\partial_{\mu} A^{\mu} = 0$.
This implies that $\Omega$ is harmonic,
$$\partial_+\partial_-\Omega = 0\eqno(17)$$
and may therefore be written as a superposition of left- and right-moving
waves.
$$\Omega(x) = \Omega^+(x^+,{\bf r}_\perp)
+ \Omega^-(x^-,{\bf r}_\perp)\eqno(18)$$
The form of the source current (16) allows introducing functions
$k^-(x^-,{\bf r}_\perp)$ and $k^+(x^+,{\bf r}_\perp)$
defined by $j_-(x^-,{\bf r}_\perp) = \partial_-k^-(x^-,{\bf r}_\perp)$
and $j_+(x^+,{\bf r}_\perp) = \partial_+ k^+(x^+,{\bf r}_\perp)$.  That is,
$$j^\alpha = \epsilon^{\alpha\beta}\partial_\beta k\eqno(19)$$
where
$$k(x) = k^+(x^+,{\bf r}_\perp) - k^-(x^-,{\bf r}_\perp)\eqno(20)$$
Note that writing $j$ in the form (19)
insures current conservation, $\partial_\alpha j^\alpha = 0$.

The electromagnetic Lagrange density in the presence of an external
current is given in the eikonal limit by
$${\cal L} = -{1 \over 4} F_{\mu\nu}F^{\mu\nu} - j^{\mu} A_{\mu}
 = {1 \over 2} \partial_i \partial_\alpha \Omega \partial_i \partial^
\alpha \Omega - j^\alpha\partial_\alpha \Omega \eqno(21)$$
As a consequence of Landau gauge (17) and current conservation,
this is a total $x^\alpha$
derivative.  Introducing the decomposition into left- and right-moving
waves, ${\cal L}$ may be written
$$\eqalign{{\cal L} &= - {1 \over 2} \partial_- \Omega^- \nabla^2 \partial_+
\Omega^+
   - {1 \over 2} \partial_+ \Omega^+ \nabla^2 \partial_- \Omega^-
   - \partial_+ k^+ \partial_- \Omega^- - \partial_- k^- \partial_+ \Omega^+
\cr
&= - \partial_-\bigl[{1\over2}\Omega^-\nabla^2\partial_+\Omega^+ +
\partial_+ k^+\Omega_-\bigr] - \partial_+\bigl[{1\over 2} \Omega^+ \nabla^2
\partial_-\Omega^- +  \partial_- k^-\Omega^+\bigr]\cr} \eqno(22)$$
Upon integrating over the entire transverse ${\bf r}_\perp$ space
and a closed surface in the two dimensional $(t,z)$ Minkowski space, the
action is given by the boundary contribution.
$$I(\Omega,k) = \int d^4x {\cal L} = \oint d\tau \int d^2 r_\perp \bigl(
{1 \over 2} \Omega^- \nabla^2 {\dot \Omega}^+ - {1 \over 2} \Omega^+
\nabla^2 {\dot \Omega}^- + {\dot k}^+ \Omega^- - {\dot k}^- \Omega^+\bigr)
\eqno(23)$$
Here all quantities are evaluated along a closed contour $x^\alpha(\tau)$
bounding the surface in the $(t,z)$ plane.  An overdot denotes
$\tau$-differenti
ation.
The equations of motion,
$$\nabla^2 {\dot \Omega}^+ = - {\dot k}^+ \qquad\qquad
  \nabla^2 {\dot \Omega}^- = - {\dot k}^-\eqno(24)$$
may be immediately integrated.
$$\eqalign{\Omega^+(x^+,{\bf r_\perp}) &= - {1 \over \nabla^2}
k^+(x^+,{\bf r_\perp}) \cr
  \Omega^-(x^-,{\bf r_\perp}) &= - {1 \over \nabla^2}
k^-(x^-,{\bf r_\perp})\cr}\eqno(25)$$
It is readily verified that this solution also follows from the full Maxwell
equations $\partial_\mu F^{\mu\nu} = j^\nu$
with the eikonal form (16) for the source
current.

Because the action (23) is first order in $\tau$, its symplectic structure
leads to the quantum commutator$^6$
$$\bigl[\Omega^+\bigl(x^+(\tau),{\bf r}_\perp\bigr),
        \Omega^-\bigl(x^-(\tau),{\bf r}_\perp^\prime\bigr)\bigr]
= {i \over 2 \pi} {\rm ln} \vert {\bf r}_\perp - {\bf r}_\perp^\prime \vert
\eqno(26)$$

To rederive the scattering amplitude (13) in this formalism, one must
calculate the $S$-matrix for the scattering process by computing the
expectation
value of the operator
$$ V=\exp i\oint d\tau\int d^2 r_\perp \left(\dot k^+\Omega^- - \dot
k^-\Omega^+ \right)$$
The relevant functional integral, corresponding to an integration over the unobserved
electromagnetic degrees of freedom, is given by
$$
\ll V\rr = {\int{\cal D}\Omega \exp iI(\Omega,k) \over\int{\cal D}\Omega \exp iI(\Omega,0)}
\eqno(27)$$
where $I(\Omega,0)$ is the
kinetic part of the action (23). Since the latter is quadratic, the functional
integral evaluates to its classical saddle-point value.  Substituting (25) in
(23) leaves
$$\eqalign{ \ll V\rr &= \exp\Biggl[ {i\over 4\pi} \oint d\tau \int d^2 r_\perp
\int d^2 r'_\perp \biggl[ k^+ (\tau,{\bf r}_\perp) \ln \left| {\bf r}_\perp -
{\bf r}'_\perp\right| \dot k^- (\tau,{\bf r}'_\perp) \cr
&\qquad\qquad - k^- (\tau,{\bf r}_\perp) \ln \left| {\bf r}_\perp - {\bf
r}'_\perp \right| \dot k^+ (\tau,{\bf r}'_\perp) \biggr] \Biggr]
\cr}\eqno(28)$$

For the two particle scattering problem, we take
$$k=e\theta(x^+-x^{(1)+})\delta({\bf r}_\perp-{\bf
r}_\perp^{(1)})+e'\theta(x^--x^{(2)-})
\delta({\bf r}_\perp-{\bf r}_\perp^{(2)})\eqno(29)$$
which corresponds to the current for a right and a left moving charged
particle, with an impact parameter ${\bf r}_\perp^{(1)}-{\bf r}_\perp^{(2)}$.
Inserting the above expression for $k$ into (28) and performing the three
integrations yields the scattering matrix.
$$S_{12}=\exp\left[i\alpha\ln\left| {\bf r}^{(1)}_\perp - {\bf
r}^{(2)}_\perp\right|^2\right]\eqno(30)$$

To see explicitly that this scattering matrix leads to
the conditions (11), it is sufficient to rewrite $S_{12}$ in terms
of the total momentum and angular momentum of the system.
The former is, of course,
$$P^\mu \equiv p^\mu_1 + p^\mu_2\eqno(31)$$
An expression for the latter is constructed from the
four-vector$^3$
$$J_\mu \equiv - {1\over \sqrt{P^2}} \epsilon_{\mu\alpha\beta\gamma}
p^\alpha_1 p^\beta_2 \left( x_1 - x_2\right)^\gamma \eqno(32\hbox{a})$$
which in the center-of-mass frame $({\bf p}_1 = - {\bf p}_2)$ has no time
component, and its space component is the total angular momentum.
$${\bf J} = {\bf L} \equiv
 {\bf r}_1\times {\bf p}_1 + {\bf r}_2 \times {\bf p}_2
\eqno(32\hbox{b})$$
Hence, ${\bf L}^2 = - J^2$, and for $p^2_1= p^2_2=0$, $S_{12}$ may be written
as
$$S_{12} = \left( -{J^2\over P^2/4}\right)^{i\alpha} \eqno(33)$$
which is the gravitational result,$^3$
with $-Gs$ replaced by $\alpha$.
We see that for free partial waves the action of the scattering matrix is
equivalent to (11), and so conclude that $S_{12}$ is
exactly equivalent to the scattering amplitude (13).

It is known that the eikonal approximation gives the correct high-energy
($s\to\infty$, $t/s\to0$) behavior of perturbative graphs when vector mesons
are exchanged, but it fails with scalar meson exchange because there the
eikonal
contributions do not dominate non-eikonal effects.$^7$  With vector exchange,
one-meson emission vertices carry an additional factor of $\sqrt{s}$ and this
serves to enhance the eikonal contribution over the the non-eikonal.  Thus one
may expect that also with graviton exchange, where the single graviton
emission vertex is enhanced by $s$, the eikonal contributions should be
dominant; in particular, the (non-renormalizable) infinities are formally
sub-dominant at large $s$.  While these remarks support the reliability of
't~Hooft's formula, an explicit check would be welcome.$^8$  Finally, we
mention that the approximations in the eikonal approach to
quantum electrodynamics are closely
related to those used in deriving the
low-energy behavior of photons.  This is similarly true for gravitons;
exact photon low-energy theorems may be extended to the
gravitational case,$^9$ where
they enjoy a universal validity, similar to 't~Hooft's scattering amplitude.

The eikonal approximation is also valid for non-Abelian vector meson exchange,
where it produces essentially Abelian results.$^{10}$  In the context of the
present investigation, this is seen by taking the non-Abelian single-particle
current to be
$${\cal J}^\mu(x) = Q(t) j^\mu(x) \eqno(34)$$
where ${\cal J}^\mu$ and $Q$ are in the Lie algebra; $j^\mu$ also depends on
the particle path $x^\mu(t)$ and is conserved, as in (1) and (6).  In order
that ${\cal J}^\mu$ be covariantly conserved, $Q$ must satisfy
$${d\over dt} Q(t)+\left[ \dot x^\mu(t) A_\mu\left(x(t)\right), Q(t)\right] =
0 \eqno(35)$$
When the eikonal {\it Ansatz\/} is made
$$E^z = 0\ \ ,\qquad B^z=0\ \ ,\qquad A_\pm = g^{-1} \partial_\pm g\ \
,\qquad {\bf A}^\perp = 0\eqno(36)$$
one finds that
$$Q(t) = g^{-1}Q_0g \eqno(37)$$
where $g$ is a group element and $Q_0$ is constant.  It is then easily seen
that extending the previous eikonal argument to the non-Abelian case amounts
to an
embedding of the Abelian results in the $Q_0$ ``direction.''  It remains an
open question whether an essentially non-Abelian eikonal approximation can be
formulated.
\medskip
\centerline{\hbox to 2in{\hrulefill}}
\smallskip
\centerline{ Professor T.~D.~Lee encouraged us to develop
these ideas and we thank him.}
\vfill
\eject
\centerline{\bf REFERENCES}
\medskip
\item{1.}G.~'t~Hooft, {\it Phys. Lett.\/} {\bf B198} (1987) 61.
\medskip
\item{2.}P.~Aichelburg and R. Sexl, {\it Gen. Rev. Grav.\/} {\bf 2} (1971)
303; T. Dray and G.~'t~Hooft, {\it Nucl. Phys.\/} {\bf B253} (1985) 173.
\medskip
\item{3.}H. Verlinde and E. Verlinde, Princeton preprint PUPT--1279 (September
1991).
\medskip
\item{4.}Aichelburg and Sexl, Ref.~[2], also discuss this limit.
\medskip
\item{5.}See for example, H. Fried, {\it Functional Methods and Models in
Quantum Field Theory\/} (MIT Press, Cambridge, MA, 1972).  In transcribing
Fried's Eq.~(9.13) into our (14), the following changes must be made.  He deals
with fermions whose electromagnetic vertex $\bar{u} (p_i) \gamma^\mu u (p_f)$
is approximated by $p^\mu/m$, $p^\mu_i \approx p^\mu_f\sim p^\mu$; we
consider bosons with vertex $p^\mu_i + p^\mu_f\sim 2p^\mu$.  Consequently, his
amplitude carries a factor $p\cdot p'/m^2\sim s/2m^2$, while the corresponding
factor in ours should be $4p\cdot p'\sim 2s$; thus one must multiply (9.13) by
$4m^2$.  Also he omits the standard kinematical factor, which for bosons is
$(2\pi)^{-2} \left( 2E_i E'_i 2E_f 2E'_f\right)^{-1/2}\sim 1/8\pi^2s$.
Therefore a factor of $m^2/2\pi^2 s$ converts Fried's amplitude (9.13)
to ours and also his $\gamma(s)$ is set to 1, for large $s$.
\medskip
\item{6.}L. Faddeev and R. Jackiw, {\it Phys. Rev. Lett.\/} {\bf 60} (1988)
1692.
\medskip
\item{7.}G. Tiktopoulos and S. Trieman, {\it Phys. Rev. D\/} {\bf 3} (1971)
1037; E. Eichten and R. Jackiw, {\it Phys. Rev. D\/} {\bf 4} (1971) 439.
\medskip
\item{8.}D. Kabat and M. Ortiz, in preparation.
\medskip
\item{9.}D. Gross and R. Jackiw, {\it Phys. Rev.\/} {\bf 166} (1968) 1287; R.
Jackiw and L. Soloviev, {\it Phys. Rev.\/} {\bf 173} (1968) 1485.
\medskip
\item{10.}J.Cornwall and G.Tiktopoulos, {\it Phys. Rev. D\/} {\bf 15} (1977)
2937.
\par
\vfill
\end